\documentclass[aps,pre,onecolumn,amsmath,amssymb]{revtex4}
\usepackage{bm}
\usepackage{graphicx}

\newcommand{\be}{\begin{eqnarray}}
\newcommand{\ee}{\end{eqnarray}}

\newcommand{\ba}{\begin{array}}
\newcommand{\ea}{\end{array}}

\begin{document}

\title{Comment on: Disproof of Bell's Theorem by Clifford Algebra Valued Local Variables}

\author{Marcin Paw\l owski}

\address{ Institute of Theoretical Physics and Astrophysics, Uniwersytet Gda\'{n}ski, PL-80-952, Gda\'{n}sk
\\
Katedra Fizyki Teoretycznej i Informatyki Kwantowej, Politechnika Gda\'{n}ska, PL-80-952, Gda\'{n}sk}

\begin{abstract}
We ''save'' Bell's Theorem by showing a flaw in Christian's argument.
\end{abstract}

\maketitle

In recent paper Christian \cite{Joy} argues that Bell's Theorem can be proved wrong by the use of Clifford Algebra valued
local realistic variables. There is an important flaw in that paper which we are to unveil and ''save'' Bell's Theorem.
Namely, the statement that: the equation (7) from that paper works out to be the equation (19) is not true. Let us rewrite
equation (7)
\be\label{7}
\mathcal{E}_{h.v.}({\bf a},{\bf b})=\int_\Lambda A_{\bf a} (\lambda)B_{\bf b}(\lambda)d\rho(\lambda)
\ee
where the observables $A_{\bf a}(\lambda)$ and $B_{\bf b}(\lambda)$ can have values $\pm1$. Now in
\be\label{19}
\mathcal{E}_{c.v.}({\bf a},{\bf b})=\int_{\mathcal{V}_3} ({\bf \mu}\cdot {\bf a} )({\bf \mu}\cdot {\bf b} )  d\rho({\bf\mu})
\ee
expressions ${\bf \mu}\cdot {\bf a} $ and ${\bf \mu}\cdot {\bf b} $ correspond to $A_{\bf a}(\lambda)$ and $B_{\bf b}(\lambda)$
respectively, but they are not equal. As Christian mentions ${\bf \mu}\cdot {\bf a} $ and ${\bf \mu}\cdot {\bf b} $ are
bivectors not scalars like $A_{\bf a}(\lambda)$ and $B_{\bf b}(\lambda)$ and since the proof of Bell's Theorem requires scalars, the
equation (19) is irrelevant to this proof. The flaw in the equation (19) comes form confusion between Clifford Algebra valued hidden variables which are acceptable,
and Clifford Algebra valued observables which are not if we are to get a scalar in the RHS of the CHSH inequality. QED.

\

This work is part of EU 6FP programme QAP.

\end{document}